\documentclass[prd,preprint,tightenlines,%
superscriptaddress,showpacs,floatfix,nofootinbib
]{revtex4}
\usepackage{epsfig}

\begin{document}
\title{Magnetic moment of the Roper resonance}
\author{T.~Bauer}
\affiliation{Institut f\"ur Kernphysik, Johannes
Gutenberg-Universit\"at, D-55099 Mainz, Germany}
\author{J.~Gegelia}
\affiliation{Institut f\"ur Kernphysik, Johannes
Gutenberg-Universit\"at,  D-55099 Mainz, Germany}
\affiliation{Institut f\"ur Theoretische Physik II, Ruhr-Universit\"at Bochum, 44780 Bochum, Germany}
\affiliation{High
Energy Physics Institute of TSU, 0186 Tbilisi, Georgia}
\author{S.~Scherer}
\affiliation{Institut f\"ur Kernphysik, Johannes
Gutenberg-Universit\"at, D-55099 Mainz, Germany}

\date{December 16, 2011}

\begin{abstract}

   The magnetic moment of the Roper resonance is calculated in the framework
of a low-energy effective field theory of the strong interactions.
   A systematic power-counting procedure is implemented by applying the complex-mass scheme.

\end{abstract}

\maketitle

\section{Introduction}

   Chiral perturbation theory \cite{Weinberg:1979kz,Gasser:1983yg}
provides a successful description of the Goldstone boson sector of
QCD (see, e.g., Ref.~\cite{Scherer:2009bt} for a recent review).
   A straightforward power counting, i.e. correspondence between the
loop expansion and the chiral expansion in terms of momenta and
quark masses at a fixed ratio \cite{Gasser:1983yg}, is obtained by
using dimensional regularization in combination with the modified
minimal subtraction scheme. Therefore, a systematic and controllable
improvement is possible in perturbative calculations of physical
quantities at low energies.
   The construction of a consistent power counting in effective field theories with
heavy degrees of freedom turns out to be a more complex problem.
   For example, power counting is violated in baryon chiral perturbation theory
if dimensional regularization and the minimal subtraction scheme are
applied \cite{Gasser:1987rb}.
   The problem has been handled by employing the heavy-baryon approach \cite{Jenkins:1990jv}
and, alternatively, by choosing a suitable renormalization scheme
\cite{Tang:1996ca,Becher:1999he,Gegelia:1999gf,Fuchs:2003qc}.
   Using the mass difference between the nucleon and the $\Delta(1232)$ resonance as an additional expansion
parameter, the $\Delta$ resonance can also be consistently included
in the framework of effective field theory
\cite{Hemmert:1997ye,Pascalutsa:2002pi,Bernard:2003xf,Pascalutsa:2006up,Hacker:2005fh}.
   On the other hand, the inclusion of heavier baryon resonances such as the Roper resonance
requires a non-trivial generalization.
   In this case the problem of power counting can be solved by using the complex-mass scheme (CMS)
\cite{Stuart:1990,Denner:1999gp,Denner:2006ic,Actis:2006rc,Actis:2008uh} which can be
understood as an extension of the on-mass-shell renormalization
scheme to unstable particles. In previous papers we have calculated
the pole masses and the widths of the $\rho$ meson and the Roper
resonance \cite{Djukanovic:2009zn,Djukanovic:2009gt}. In the current
paper we consider the magnetic moment of the Roper up to ${\cal
O}(q^3)$.\footnote{Here, $q$ stands for small parameters of the
theory such as the pion mass.} While the extraction of these
quantities from experimental measurements at present seems to be
unfeasible, our expression for the magnetic moment may be used in
the context of lattice QCD. Effective field theories predict the
quark-mass dependence of physical observables and can be used to
extrapolate simulations in the framework of lattice QCD performed at
unphysically large masses of the light quarks. In return, lattice
QCD provides a way to determine the low-energy constants from the
underlying theory.

\medskip

\section{Effective Lagrangian}

   In this section we specify the effective Lagrangian
relevant for the subsequent calculation of the electromagnetic
vertex of the Roper at ${\cal O}(q^3)$. We include the pion, the
nucleon, the Roper, and the $\Delta$ as explicit degrees of freedom.
The effects of other degrees of freedom are buried in low-energy
coupling constants.  We write the effective Lagrangian
as\footnote{To simplify the notation only bare masses are supplied
with a subscript $0$.}
\begin{equation}
\mathcal{L} =  \mathcal{L}_{0}+ {\cal L}_\pi+ \mathcal{L}_{R}
+\mathcal{L}_{N R}+\mathcal{L}_{\Delta R}\,, \label{lagrFull}
\end{equation}
where $\mathcal{L}_{0}$ is given by
\begin{eqnarray}
\mathcal{L}_{0} & = & \bar{N}\,(i
D\hspace{-.65em}/\hspace{.1em}-m_{N 0})N + \bar{R}(i
D\hspace{-.65em}/\hspace{.1em}-m_{R 0}) R \nonumber\\
&&
 - \bar\Psi_\mu\xi^{\frac{3}{2}}
\biggl[(i {D\hspace{-.65em}/\hspace{.1em}}-m_{\Delta 0})\,g^{\mu\nu}
-i\,(\gamma^{\mu}D^{\nu}+\gamma^{\nu}D^{\mu})+i\,\gamma^{\mu}
{D\hspace{-.65em}/\hspace{.3em}}\gamma^{\nu} + m_{\Delta
0}\,\gamma^{\mu}\gamma^{\nu}\biggr]\xi^{\frac{3}{2}} \Psi_\nu .
\end{eqnarray}
   Here, $N$ and $R$ denote nucleon and Roper isospin doublets
with bare masses $m_{N 0}$ and $m_{R 0}$, respectively.
   $\Psi_\nu$ represents the vector-spinor isovector-isospinor
Rarita-Schwinger field of the $\Delta$ resonance
\cite{Rarita:1941mf} with bare mass $m_{\Delta 0}$,
$\xi^{\frac{3}{2}}$ is the isospin-$3/2$ projector (see
Ref.~\cite{Hacker:2005fh} for more details).
   The covariant derivatives are defined as follows:
\begin{eqnarray}
D_\mu H & = & \left( \partial_\mu + \Gamma_\mu-i\,v_\mu ^{(s)}\right) H\,, \nonumber\\
\left(D_\mu\Psi\right)_{\nu,i} & = &
\partial_\mu\Psi_{\nu,i}-2\,i\,\epsilon_{ijk}\Gamma_{\mu,k} \Psi_{\nu,j}+\Gamma_\mu\Psi_{\nu,i}
-i\,v_\mu^{(s)}\Psi_{\nu,i}\,,\nonumber\\
\Gamma_\mu & = &
\frac{1}{2}\,\left[u^\dagger \partial_\mu u +u
\partial_\mu u^\dagger-i\,\left( u^\dagger v_\mu u+u v_\mu u^\dagger
\right)\right]=\tau_k\Gamma_{\mu,k}, \label{cders}
\end{eqnarray}
where $H$ stands either for the nucleon or the Roper.
   The pion fields are contained in the unimodular, unitary, $(2\times 2)$
matrix $U$ and $u=\sqrt{U}$.
   The external electromagnetic four-vector potential ${\cal A}_\mu$
enters into $v_\mu= - e\,\frac{\tau_3}{2}\,{\cal A}_\mu$
and $v_\mu^{(s)}= - \frac{e}{2}\,{\cal A}_\mu$ ($e^2/(4\pi)\approx 1/137, e>0$).

   The lowest-order Goldstone-boson Lagrangian including the quark-mass term and
the interaction with the external electromagnetic four-vector
potential ${\cal A}_\mu$ reads
\begin{equation}
\label{l2} {\cal L}_\pi^{(2)} =
\frac{F^2}{4}\mbox{Tr}\left(\partial_\mu U \partial^\mu
U^\dagger\right) +\frac{F^2 M^2}{4}\mbox{Tr} \left( U^\dagger+ U
\right)+i\,\frac{F^2}{2}\mbox{Tr}\left[\left(\partial_\mu U
U^\dagger +
\partial_\mu U^\dagger U\right) v^\mu \right]\,.
\end{equation}
   $F$ denotes the pion-decay constant in the chiral limit:
$F_\pi=F[1+{\cal O}(q^2)]=92.4$ MeV; $M$ is the pion mass at leading
order in the quark-mass expansion: $M^2=2 B\hat m$, where $B$ is
related to the quark condensate $\langle \bar q q\rangle_0$ in the
chiral limit \cite{Gasser:1983yg}.

   The interaction terms $\mathcal{L}_{R}$, $\mathcal{L}_{NR}$, and $\mathcal{L}_{\Delta R}$
are constructed in analogy to Ref.~\cite{Borasoy:2006fk}.
   The leading-order (${\cal O}(q)$) pion-Roper coupling is given by
\begin{equation}
{\cal L}_R^{(1)} = \frac{g_R}{2}\,\bar R \gamma^\mu\gamma_5 u_\mu R\,,
\label{Rpiint1}
\end{equation}
where $g_R$ is an unknown coupling constant and
\begin{equation}
u_\mu =i \left[u^\dagger \partial_\mu u -u \partial_\mu u^\dagger
-i\,\left( u^\dagger v_\mu u-u v_\mu u^\dagger\right)\right].
\label{umudef}
\end{equation}
   The second- and third-order Roper Lagrangians relevant for our
calculation read
\begin{eqnarray}
{\cal L}_R^{(2)} & = & \bar R\left[  \frac{c_{6}^*}{2}
\,f^+_{\mu\nu}+\frac{c_{7}^*}{2} \,v^{(s)}_{\mu\nu}
\right]\,\sigma^{\mu\nu}R+\cdots \,,\nonumber\\
{\cal L}_R^{(3)} & = & \frac{i}{2}\,d_{6}^* \bar R\left[
D^\mu,f^+_{\mu\nu}\right]\,D^\nu R+{\rm H.c.}+ 2\,i\,d_{7}^* \bar
R\left(
\partial^\mu v^{(s)}_{\mu\nu}\right)\,D^\nu R+{\rm H.c.}+ \cdots \,,
\label{Rpiint2}
\end{eqnarray}
where
\begin{eqnarray}
v_{\mu\nu}^{(s)} & = & \partial_\mu v^{(s)}_\nu - \partial_\nu v^{(s)}_\mu\,,\nonumber\\
f_{\mu\nu}^{+} & = & u f_{\mu\nu} u^\dagger +u^\dagger f_{\mu\nu} u\,,\nonumber\\
f_{\mu\nu} & = & \partial_\mu v_\nu - \partial_\nu v_\mu-i
[v_\mu,v_\nu], \label{bbks}
\end{eqnarray}
and $c_{6}^*$, $c_{7}^*$, $d_{6}^*$, and $d_{7}^*$ are unknown
coupling constants.
   The ellipsis denote those terms of the most
general second- and third-order Roper Lagrangians which do not
contribute to the electromagnetic vertex of the Roper at ${\cal
O}(q^3)$ and H.c.~refers to the Hermitian conjugate.
   The leading-order interaction between the nucleon and the Roper is given by
\begin{equation}
{\cal L}_{N R}^{(1)} = \frac{g_{N R}}{2}\,\bar R
\gamma^\mu\gamma_5 u_\mu N+ {\rm H.c.}\, \label{Rpiint11}
\end{equation}
with an unknown coupling constant $g_{N R}$.
   Finally, the leading-order interaction between the $\Delta$ and the Roper
reads
\begin{equation}
{\cal L}_{\Delta R}^{(1)}= - g_{\Delta R} \,\bar{\Psi}_{\mu}
\,\xi^{\frac{3}{2}} \,(g^{\mu\nu}
+\tilde{z}\,\gamma^{\mu}\gamma^{\nu})\, u_{\nu}\, R + {\rm H.c.}\,,
\label{pND}
\end{equation}
where $g_{\Delta R}$ is a coupling constant and we take the
''off-mass-shell parameter''  $\tilde z=-1$. Note that at ${\cal
O}(q^3)$ the $N\Delta$ Lagrangian does not contribute.

\section{Perturbation theory, renormalization, and power counting}

The CMS
\cite{Stuart:1990,Denner:1999gp,Denner:2006ic,Actis:2006rc,Actis:2008uh}
originates from the Standard Model where it was developed to derive
properties of $W$, $Z_0$, and Higgs bosons obtained from resonant
processes.
   What makes the situation somewhat different in the case of the strong interactions
is the fact that hadrons, including resonances, are thought to be composite objects
made of quarks and gluons.
   The characteristic properties of hadron resonances eventually have to
be described by QCD.
   Within the present effective-field-theory approach, to a given resonance
we assign an explicit field with corresponding spin, isospin, and parity content.
   Furthermore, for a generic resonance $R$, we introduce a complex renormalized mass
$z_\chi$ defined as the location of the corresponding complex pole position in the
chiral limit, $z_\chi=m_{R\chi}-i\Gamma_{R\chi}/2$.
   We assume $\Gamma_{R\chi}$ to be small in comparison to both $m_{R\chi}$ and
the scale of spontaneous chiral symmetry breaking, $\Lambda_\chi=4\pi F$.
   Corrections to the complex pole position due to the finite quark masses are treated
perturbatively.
   Our perturbative approach to EFT is based on the path integral formalism.
   In this framework the physical quantities are obtained from Green's functions
represented by functional integrals.
   The integration over classical fields corresponding to particles is performed in the
standard way, i.e., the Gaussian part is treated non-perturbatively and the rest
perturbatively.
   In particular, the functional integral is performed for both stable and unstable
degrees of freedom.
   For stable particles the path integral formalism is equivalent to the operator
formalism based on the Dirac interaction representation.
   Unfortunately, it is not obvious how to apply this representation
to field operators for unstable particles, because, strictly speaking, there is no
free Hamiltonian for unstable particles.
   Therefore, we stick to the functional integral where one can perform the integration
independently of the nature of the field.

   In the following, we apply the CMS to have a consistent power counting also applicable
to loop diagrams.
   This renormalization scheme is realized by splitting the bare parameters
(and fields) of the Lagrangian into, in general, complex
renormalized parameters and counter terms. We choose the renormalized
masses as the poles of the dressed propagators in the chiral limit:
\begin{eqnarray}
m_{R 0} & = & z_\chi+\delta z_\chi\,,\nonumber \\
m_{N 0} & = & m_\chi+\delta m \,,\nonumber\\
m_{\Delta 0} & = & z_{\Delta \chi} + \delta z_{\Delta \chi} \,,
\label{barerensplit}
\end{eqnarray}
where $z_{\chi}$ is the complex pole of the Roper propagator in the
chiral limit, $m_\chi$ is the mass of the nucleon in the chiral
limit, and $z_{\Delta \chi}$ is the complex pole of the $\Delta$
propagator in the chiral limit.    We include the renormalized
parameters $z_\chi$, $m$, and $z_{\Delta \chi}$ in the propagators
and treat the counter terms perturbatively.
   The renormalized couplings $c_{6}^*$ and $c_{7}^*$ of ${\cal L}_R^{(2)}$ are chosen such that the corresponding
counter terms exactly cancel the power-counting-violating parts of
the loop diagrams.

While the starting point is a Hermitian Lagrangian in terms of bare
parameters and fields, the CMS involves complex parameters in the
basic Lagrangian and complex counter terms. Although the application
of the CMS seems to violate unitarity, the bare Lagrangian is
unchanged and unitarity cannot be violated in the complete theory.
On the other hand, it is not obvious that the approximate
expressions to the $S$-matrix generated by perturbation theory also
satisfy the unitarity condition since the conventional Cutkosky
cutting equations \cite{Cutkosky:1960sp} are not valid in the
framework of CMS. However, it is possible to derive generalized
cutting rules for loop integrals involving propagators with complex
masses to show that unitarity is satisfied perturbatively
\cite{BGJS}. In agreement with Ref. \cite{Veltman:1963th}, the
$S$-matrix connecting stable states only is unitary.

   We organize our perturbative calculation by applying the standard
power counting of Refs.~\cite{Weinberg:1991um,Ecker:1995gg} to the
renormalized diagrams, i.e., an interaction vertex obtained from
an ${\cal O}(q^n)$ Lagrangian counts as order $q^n$, a pion
propagator as order $q^{-2}$, a nucleon propagator as order
$q^{-1}$, and the integration of a loop as order $q^4$.
   In addition, we assign the order $q^{-1}$ to the
$\Delta$ propagator and to the Roper propagator.
   Within the CMS, such a power counting is respected by the renormalized loop diagrams in the
range of energies close to the Roper mass.
   In practice, we implement this scheme by subtracting the loop diagrams at complex ''on-mass-shell''
points in the chiral limit.

   { When calculating an observable, we do not perform an expansion in powers of the mass
differences between the Roper and the nucleon or the Roper and the $\Delta$.
   Rather we calculate  the chiral corrections to the magnetic moment of the Roper as a
series in powers of the pion mass which is either divided by large scales, like $4 \pi F$ and the heavy masses,
or multiplied by coupling constants which contain (inverse powers of) hidden
large scales.
   As the omitted neighboring resonances, like $N(1535)$, couple weakly to the Roper
resonance,
inverse powers of small scales (mass differences between the Roper and the omitted resonances) which are hidden in
low-energy coupling constants of our effective theory
are enhanced by inverse powers of small couplings (corresponding to the weak coupling of the Roper resonance to its neighbors)
and therefore effectively appear as large scales.}

\medskip

   The dressed propagator of the Roper can be written as
\begin{equation}
i S_R(p) = \frac{i}{p\hspace{-.45 em}/\hspace{.1em}-z_\chi
-\Sigma_R(p\hspace{-.45 em}/\hspace{.1em})}\,,\label{dressedDpr}
\end{equation}
where $-i\Sigma_R (p\hspace{-.45 em}/\hspace{.1em})$ denotes the sum
of one-particle-irreducible diagrams contributing to the Roper
two-point function.
   The pole of the dressed propagator $S_R$ is obtained by solving the equation
\begin{equation}
z - z_\chi -\Sigma_R(z)=0\,. \label{poleequation}
\end{equation}
   We define the pole mass and the width as the real part and $(-2)$ times
the imaginary part of the pole \cite{Djukanovic:2007bw}, respectively,
\begin{equation}
z = m_R -i\,\frac{\Gamma_R}{2} \,.\label{poleparameterized}
\end{equation}
 Some of the phenomenological analyses and dynamical models describe the Roper
as a double-pole structure (see, e.g., Refs.\ \cite{Arndt:1985vj,Cutkosky:1990zh}).
   As the self energy in Eq.~(\ref{poleequation}) is a multi-valued function, one might
be tempted to look for several solutions of this equation.
   Although the numbering of sheets is a matter of convention, it is our understanding
that in the standard nomenclature only poles on the second sheet are relevant for the
physical amplitude and should be interpreted as resonances.
   Within our perturbative approach, Eq.~(\ref{poleequation}) has a unique solution
on the second sheet.
   This solution is obtained as a power series in terms of the expansion parameter(s)
of the perturbation theory.

\medskip

\begin{figure}
\epsfig{file=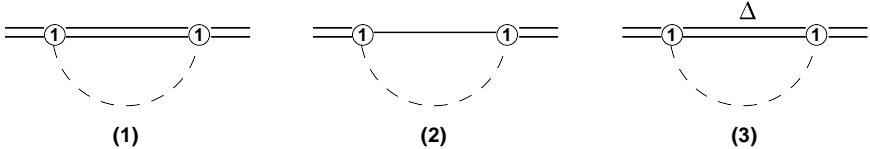,width=0.7\textwidth}
\caption[]{\label{DeltaMassInd:fig} One-loop self-energy diagrams of
the Roper. Dashed and solid lines refer to the pion and nucleon,
respectively, and double-solid lines correspond to the Roper and
delta. The numbers in the vertices indicate the chiral order.}
\end{figure}

\noindent Close to the pole the Roper propagator can be
parameterized as
\begin{equation}
i S_R(p) = \frac{i\, Z_R}{p\hspace{-.45 em}/\hspace{.1em}-z}+ {\rm
n.p.}\,.\label{dressedDpr2}
\end{equation}
    The residue $Z_R$ (wave function renormalization constant of the
Roper) is a complex-valued quantity and n.p. stands for the non-pole
part. This is in full agreement with Ref.~\cite{Gegelia:2009py},
where we have shown that physical quantities characterizing unstable
particles have to be extracted at pole positions using
complex-valued wave function renormalization constants. Up to ${\cal
O}(q^3)$, $Z_R$ is obtained by calculating the Roper self-energy
diagrams shown in Fig.~\ref{DeltaMassInd:fig}. We do not give its
explicit expression here.

\begin{figure}
\epsfig{file=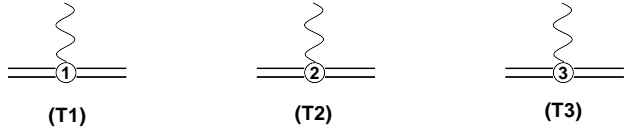,width=0.5\textwidth} \caption[]{
\label{roperFFTree:fig} Tree diagrams contributing to the elastic
electromagnetic form factors of the Roper resonance. Double-solid
and wiggly lines correspond to the Roper and external
electromagnetic source, respectively. The numbers in the vertices
indicate the chiral order.}
\end{figure}

\begin{figure}
\epsfig{file=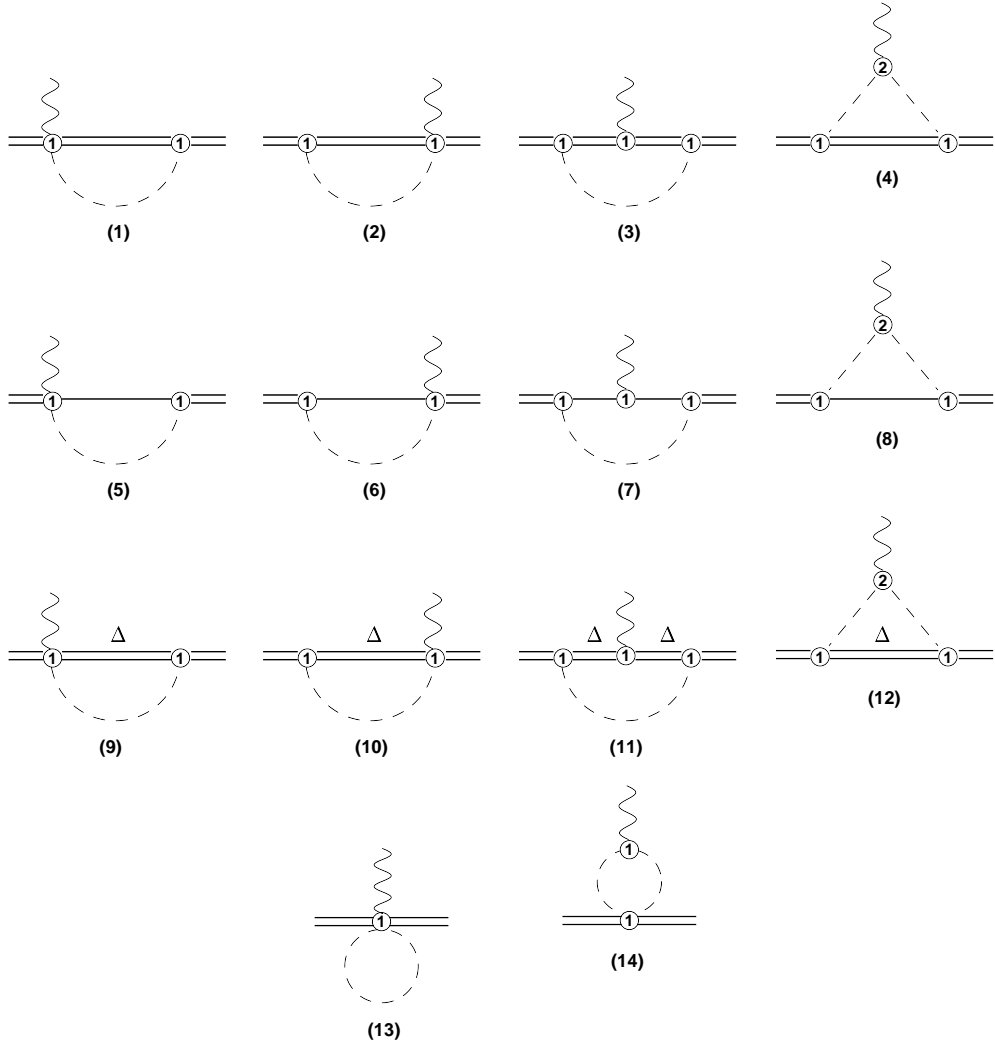,width=0.8\textwidth}
\caption[]{ \label{MassWidth:fig} Loop diagrams contributing to the
elastic electromagnetic form factors of the Roper resonance. Dashed,
wiggly, and solid lines correspond to pion, nucleon, and external
electromagnetic source, respectively; double-solid lines correspond
to the Roper and $\Delta$. The numbers in the vertices indicate the
chiral order.}
\end{figure}

\medskip

\section{Magnetic moment}
   Using Lorentz covariance and the discrete symmetries,
the most general electromagnetic vertex of a spin-1/2 field may be parameterized
in terms of 12 Dirac structures multiplied by form functions depending on three scalar
variables, e.g., $p_i^2$, $p_f^2$, and $q^2$, where $q=p_f-p_i$ \cite{Bincer:1959tz,Naus:1987kv,Koch:2001ii}.
   For charged fields, the Ward-Takahashi \cite{Ward:1950xp,Takahashi:1957xn} identity
provides certain constraints among the form functions.
   For a stable particle such as the nucleon, on-shell kinematics corresponds to $p_i^2=p_f^2=m^2_N$,
and the form functions reduce to conventional form factors of $q^2$, say, Dirac and
Pauli form factors $F_1$ and $F_2$, respectively.
   For unstable particles such as the Roper resonance, the analogous kinematical point is given by the
pole position, i.e., $p_i^2=p_f^2=z^2$.
   In Ref.~\cite{Gegelia:2009py} we described a method how to extract from the general vertex
only those pieces which survive at the pole.
   To that end, we introduced ''Dirac spinors'' $\bar w^i$ and $w^j$ with complex masses $z$ which
essentially correspond to half of the projection operators $\Lambda_+=\sum_j w^j \bar w^j$
used in Refs.\ \cite{Naus:1987kv,Koch:2001ii} for the initial and final lines.
   In terms of these ''Dirac spinors,''
the renormalized vertex function for
$p_f^2=p_i^2=z^2$ may be written in terms of two form factors,
\begin{equation}
\sqrt{Z_R}\,\bar w^i(p_f) \Gamma^\mu(p_f,p_i) w^j(p_i)\sqrt{Z_R}=
\bar w^i(p_f)\left[\,\gamma^\mu\, F_1(q^2)  + \frac{i
\,\sigma^{\mu\nu}\, q_\nu}{2\,m_N}\,F_2(q^2)\,\right]w^j(p_i) \,,
\label{FF}
\end{equation}
where  $m_N$ is the physical mass of the nucleon.\footnote{Note the
different normalization of the magnetic form factor.}
   Both electromagnetic form factors of the Roper are complex-valued functions
even for $q^2<0$ because of the resonance character of the Roper.
   As in the case of an on-shell nucleon,
the third form function vanishes at the pole because of
current conservation or time-reversal invariance.

    To ${\cal O}(q^3)$, the vertex function $\Gamma^\mu(p_f,p_i)$
obtains contributions from three tree diagrams (see
Fig.~\ref{roperFFTree:fig}) and fourteen loop diagrams (see
Fig.~\ref{MassWidth:fig}).
   By multiplying the tree-order contribution with the wave function renormalization
constant, one subtracts all power-counting-violating contributions of loop
diagrams to the $F_1$ form factor.
   We obtain $F_1(0)=(1+\tau_3)/2$ in agreement with the Ward identity.
   This means that, as expected, the electric charge of the Roper does
not receive any strong corrections.
   On the other hand, the loop contributions to
the magnetic form factor contain power-counting-violating terms.
These parts  are analytic in the squared pion mass and momenta. They
are subtracted from the loop diagrams and absorbed in the
renormalization of the couplings $c_{6}^*$ and $c_7^*$.

    The anomalous magnetic moment in units of the nuclear magneton is defined as
\begin{equation}
\kappa_R = F_2(0). \label{RMM}
\end{equation}
   Since both the four-momentum $q^\mu$ as well as the polarization vector $\epsilon^\mu$
count as ${\mathcal O}(q)$ our calculation yields the magnetic
moment to ${\mathcal O}(q)$.
   The tree-order result for $\kappa_R$ is given by
\begin{equation}
\kappa_R^{\rm tree} = 2\,m_N
\left(\frac{c_7^*}{2}+\tau_3\,c_6^*\right).
\end{equation}
   In order to show that the subtracted loop contributions satisfy the power
counting we divide the diagrams of Fig.~\ref{MassWidth:fig} into three separate
classes.
   Diagrams potentially violating the power counting are loop
diagrams with internal Roper, nucleon, and delta lines
which we refer to as classes A, B, and C, respectively.
   We denote the respective contributions to the magnetic moment by
$\kappa^{A,B,C}_{R}$.

   At first, we consider the contribution of $\kappa^{A}_{R}$.
   Dividing the expression by $M$ and taking the limit $M\rightarrow0$ yields
\begin{equation}
\lim_{M\to 0}\frac{\kappa^A_{R}}{M} =-g_R^2\frac{m_N}{8 \pi
F^2 }\:\tau_3.
\end{equation}
   Replacing the low-energy constant $g_R$ with $g_A$, this expression coincides with
the non-analytic contribution to the anomalous magnetic moment of
the nucleon \cite{Gasser:1987rb,Fuchs:2003ir}.
   Next, we analyze the contributions stemming from
$\kappa^{B}_{R}$.
   For a fixed and finite mass difference $z_\chi-m_\chi$, the limit $M\rightarrow0$ is zero
\begin{equation}
\lim_{M\rightarrow0}\frac{\kappa^B_{R}}{M} =0.
\end{equation}
If $z_\chi-m_\chi$ scales as $\alpha M$ the limit $M\rightarrow0$ is
given by
\begin{equation}
\lim_{M\rightarrow0}\frac{\kappa^B_{R}}{M}
=g_{NR}^2\,\frac{m_N}{4\pi^2 F^2 }\,g\left(\alpha\right)\,\tau_3,
\end{equation}
with
\begin{equation}
g\left(\alpha\right)= i\pi\left(\sqrt{\alpha^2-1}-\alpha\right)+
\alpha \ln(2 \alpha)-\sqrt{\alpha^2-1}\ln(\alpha+\sqrt{\alpha^2-1}).
\end{equation}
Taking the limit $M\rightarrow0$ after the limit $z_\chi\rightarrow
m_\chi$ results in
\begin{equation}
\lim_{M\rightarrow0}\,\left(\lim_{m_\chi\rightarrow
z_\chi}\frac{\kappa^B_{R}}{M} \right)=-g_{NR}^2\,\frac{m_N}{8 \pi
F^2 }\tau_3.
\end{equation}
Similar results are obtained for $\kappa^{C}_{R}$. For fixed and
finite mass difference $z_\chi-z_{\Delta\chi}$ the limit
$M\rightarrow0$ yields
\begin{equation}
\lim_{M\rightarrow0}\frac{ \kappa^C_{R}}{M}=0.
\end{equation}
If $z_\chi-z_{\Delta\chi}$ scales as $\beta M$ the limit
$M\rightarrow0$ is given by
\begin{equation}
\lim_{M\rightarrow0}\frac{\kappa^C_{R}}{M} =g_{\Delta
R}^2\,\frac{m_N}{9 \pi^2F^2 }\,g\left(\beta\right)\,\tau_3.
\end{equation}
Taking the limit $M\rightarrow0$ after $z_\chi\rightarrow
z_{\Delta\chi}$ one finds
\begin{equation}
\lim_{M\rightarrow0}\,\left(\lim_{m_\Delta\rightarrow
z_\chi}\frac{\kappa^C_{R}}{M} \right)=-g_{\Delta R}^2\,\frac{m_N}{18 \pi
F^2 }\,\tau_3.
\end{equation}
The above analysis shows that the renormalized loop diagrams satisfy
the power counting regardless of how the various mass differences
are treated.

\medskip

   To estimate the loop contributions to the anomalous magnetic moment of the Roper
we substitute \cite{PDG} $F= 0.092\,\, {\rm GeV}$, $M = 0.140\,\,
{\rm GeV}$, $m_\chi=0.940\,\, {\rm GeV}$, $z_{\Delta \chi} = (1.210
- 0.100\,i/2)\,\, {\rm GeV}$, $z_\chi = (1.365-0.190\,i/2)\,\, {\rm
GeV}$, $\mu = 1\,\, {\rm GeV}$, $g_R=1$, $g_{\Delta R} = 1$,
$g_{NR}= 0.45$ \cite{Borasoy:2006fk} and obtain
\begin{equation}
\kappa_R = (0.055+ 0.090\,i)-(0.223+ 0.156\,i)
   \,\tau_3. \label{munumeric}
\end{equation}
\begin{figure}
\epsfig{file=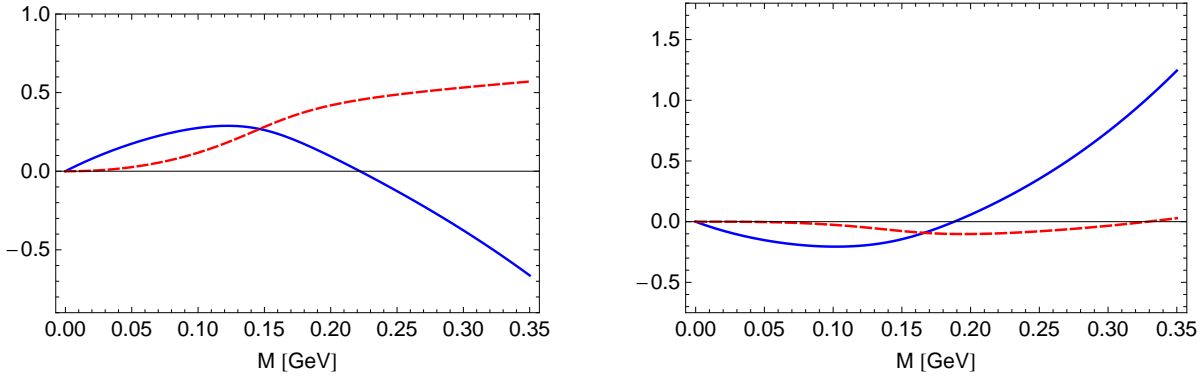,width=1.0\textwidth}
\caption[]{\label{RMM:fig} One-loop contributions to the anomalous
magnetic moment of the Roper as functions of the pion mass. The left
figure corresponds to the neutral and the right to the charged
resonances. The solid and dashed lines indicate the real and
imaginary parts, respectively.}
\end{figure}
    Figure \ref{RMM:fig} shows the loop contribution to the anomalous magnetic moment of
the Roper as a function of the lowest-order pion mass $M$, where
$M^2=2 B \hat m$ \cite{GellMann:1968rz}.

\section{Summary}
   To summarize, we have calculated the magnetic moment of the Roper
resonance up to and including order $q^3$ using effective-field-theory
techniques.
   To obtain a systematic power counting for energies
around the mass of the Roper, we applied the CMS
which is a generalization of the on-mass-shell renormalization for
unstable particles. Unrenormalized contributions of loop diagrams to
the magnetic moment contain power-counting-violating terms. However,
these terms are analytic in the squared pion mass and the momenta
and can be systematically absorbed in the renormalization of the
available low-energy coupling constants. The renormalized loop
diagrams satisfy the power counting regardless of how the Roper and
nucleon as well as the Roper and delta mass differences are treated.

   At next-to-next-to-leading order, ${\mathcal O}(q^3)$, only the
isovector anomalous magnetic moment receives a loop contribution.
   Analogously to the nucleon, the loop contribution to the isoscalar anomalous
magnetic moment {starts with order} ${\mathcal O}(q^4)$.\footnote{In manifestly Lorentz-invariant baryon
chiral perturbation theory, a calculation at ${\cal O}(q^n)$, in general, not only produces contributions of
${\cal O}(q^n)$ but also a string of higher-order terms of ${\cal O}(q^{n+i})$ with $i=1,2,3,\ldots$
\cite{Gasser:1987rb}. For the isoscalar magnetic moment, the leading term at ${\cal O}(q^3)$ vanishes and only
small contributions beyond ${\cal O}(q^3)$ survive [see Eq.~(\ref{munumeric})].}
   Due to the unstable character of the Roper, the loop contributions
to the anomalous magnetic moment feature an imaginary part which is of the
same order of magnitude as the corresponding real part.

   At present, an extraction of the elastic electromagnetic form
factors of the Roper from experimental measurements appears to be
unrealistic.
   However, our expressions for the anomalous magnetic
moment may be used in the context of lattice extrapolations.
   Moreover, lattice QCD provides for an opportunity to determine the
five unkown parameters.
   A fit of our expressions to lattice data at different values for the pion
mass results in a complete theoretical prediction of the anomalous magnetic
moments of the Roper.

\medskip

\acknowledgments

   When calculating the loop diagrams we made use of the package
FeynCalc \cite{Mertig:1990an} and computer programs written by D.
Djukanovic.
   This work was supported by the Deutsche
Forschungsgemeinschaft (SFB 443) and Georgian National
Foundation grant GNSF/ST08/4-400.
   T.~Bauer would like to thank the
German Academic Exchange Service (DAAD) for financial support.

\section{appendix}

   Making use of dimensional regularization with $n$ the number of space-time
dimensions, the loop functions are given as \cite{Denner:2005nn}
\begin{eqnarray}
A_0\left(m^2\right) &=&  \frac{(2 \pi\,\mu)^{4-n}}{i\,\pi^2}\,\int
\frac{d^nk}{k^2-m^2+i \epsilon}
=-32 \pi ^2 \lambda\,  m^2-2\,m^2 \ln \frac{m}{\mu}\,, \nonumber\\
B_0\left(p^2,m_1^2,m_2^2\right) &=& \frac{(2
\pi\,\mu)^{4-n}}{i\,\pi^2} \, \int\frac{d^nk
}{\left[k^2-m_1^2+i \epsilon\right]\left[(p+k)^2-m_2^2+i \epsilon\right]}\nonumber\\
& = &
-32 \pi ^2 \lambda +2 \ln \frac{\mu}{m_2}-1 - \frac{\omega}{2} \ {}_2F_1\left(1,2;3;\omega\right)\nonumber\\
&&  -\frac{1}
{2}\left(1+\frac{m_2^2}{m_1^2(\omega-1)}\right)\,{}_2F_1\left(1,2;3;1+\frac{m_2^2}{m_1^2(\omega-1)}\right)
\,,\nonumber\\
\omega & = & \frac{m_1^2
-m_2^2+p^2+\sqrt{\left(m_1^2-m_2^2+p^2\right)^2 -4m_1^2 p^2}}{2
m_1^2}\,, \label{oneandtwoPF}
\end{eqnarray}
where ${}_2F_1\left(a,b;c;z\right)$ is the standard hypergeometric
function, $\mu$ is the scale parameter of the dimensional
regularization and
\begin{equation}
\lambda =
\frac{1}{16\,\pi^2}\left\{\frac{1}{n-4}-\frac{1}{2}\,\left[\ln(4
\pi)+\Gamma'(1)+1\right]\right\}\,. \label{lambdadef}
\end{equation}

\medskip

\noindent By writing
\begin{equation}
F_2(t)= m_N [\,G_1(t)+\tau_3 G_2(t)], \label{f2defG}
\end{equation}
we obtain the loop contributions as:
\begin{eqnarray}
G_1^{\rm loop}(0) & = & \frac{3 g_R^2}{16 F^2 z_\chi \left(M^2-4
   z_\chi^2\right) \pi ^2}
   \Biggl\{\left[z_\chi^2-A_0\left(z_\chi^2\right)-\left(M^2-3
   z_\chi^2\right)
   B_0\left(z_\chi^2,M^2,z_\chi^2\right)\right]
   M^2\nonumber\\
&& +\left(M^2-2 z_\chi^2\right)
   A_0\left(M^2\right)\Biggr\} + \frac{3 g_{NR}^2 (m_\chi+z_\chi)^2}{64
   F^2 m_\chi z_\chi^3 \left[(m_\chi+z_\chi)^2-M^2\right] \pi ^2}\nonumber\\
& \times &  \Biggl\{z_\chi
   A_0\left(m_\chi^2\right) M^2+m_\chi \left(m_\chi
   (m_\chi+z_\chi)-M^2\right) A_0\left(M^2\right) \nonumber\\
& - & m_\chi
   \Biggl[M^2 z_\chi^2+m_\chi (m_\chi-z_\chi)
   \left((m_\chi+z_\chi)^2-M^2\right)
   B_0\left(z_\chi^2,0,m_\chi^2\right) \nonumber\\
& + & \left(-M^4+m_\chi
   (2 m_\chi+z_\chi) M^2-m_\chi (m_\chi-z_\chi)
   (m_\chi+z_\chi)^2\right)
   B_0\left(z_\chi^2,M^2,m_\chi^2\right)\Biggr]\Biggr\}\nonumber\\
& + & \frac{g_{\Delta R}^2}{864 F^2
   z_{\Delta \chi}^4 z_\chi^3 \pi ^2} \, \Biggl\{-2 (z_{\Delta \chi}-z_\chi) \left(9
   z_{\Delta \chi}^4-14 z_\chi z_{\Delta \chi}^3+8 z_\chi^2 z_{\Delta \chi}^2+2
   z_\chi^3 z_{\Delta \chi}+z_\chi^4\right)\nonumber\\
& \times &
   B_0\left(z_\chi^2,0,z_{\Delta \chi}^2\right)
   (z_{\Delta \chi}+z_\chi)^3+M^2 z_\chi^2 \biggl[\left(9
   z_{\Delta \chi}^2 + 4 z_\chi z_{\Delta \chi}-z_\chi^2\right) M^2 \nonumber\\
& + & 4
   z_\chi \left(-6 z_{\Delta \chi}^3-8 z_\chi
   z_{\Delta \chi}^2+z_\chi^2 z_{\Delta \chi} +  4 z_\chi^3\right)\biggr]+2
   \biggl[\left(9 z_{\Delta \chi}^2+4 z_\chi
   z_{\Delta \chi}-z_\chi^2\right) M^4 \nonumber\\
& - & (z_{\Delta \chi}+z_\chi) \left(18
   z_{\Delta \chi}^3-10 z_\chi z_{\Delta \chi}^2+7 z_\chi^2 z_{\Delta \chi}-3
   z_\chi^3\right) M^2\nonumber\\
& + & (z_{\Delta \chi}+z_\chi)^2 \left(9
   z_{\Delta \chi}^4-14 z_\chi z_{\Delta \chi}^3+8 z_\chi^2 z_{\Delta \chi}^2+2
   z_\chi^3 z_{\Delta \chi}+z_\chi^4\right)\biggr] A_0\left(M^2\right)\nonumber\\
& - & 2 M^2 \left[\left(9 z_{\Delta \chi}^2+4 z_\chi
   z_{\Delta \chi}-z_\chi^2\right) M^2+2 \left(-9 z_{\Delta \chi}^4-4
   z_\chi z_{\Delta \chi}^3+6 z_\chi^2
   z_{\Delta \chi}^2+z_\chi^4\right)\right]\nonumber\\
& \times & A_0\left(z_{\Delta \chi}^2\right) -  2
\left[M^2-(z_{\Delta \chi}+z_\chi)^2\right] \biggl[\left(9 z_{\Delta
\chi}^2+4 z_\chi
   z_{\Delta \chi}-z_\chi^2\right) M^4 \nonumber\\
& - & 2 \biggl[9 z_{\Delta \chi}^4-5
   z_\chi z_{\Delta \chi}^3-10 z_\chi^2 z_{\Delta \chi}^2 + z_\chi^3
   z_{\Delta \chi}-z_\chi^4\biggr] M^2 \nonumber\\
& + & (z_{\Delta \chi}^2-z_\chi^2)
   \left(9 z_{\Delta \chi}^4-14 z_\chi
   z_{\Delta \chi}^3+8 z_\chi^2 z_{\Delta \chi}^2+2 z_\chi^3
   z_{\Delta \chi}+z_\chi^4\right)\biggr] B_0\left(z_\chi^2,M^2,z_{\Delta \chi}^2\right)\Biggr\}, \\
G_2^{\rm loop}(0) & = & \frac{g_R^2}{16 F^2
   z_\chi \left(M^2-4 z_\chi^2\right) \pi ^2}\,
\Biggl\{-A_0\left(z_\chi^2\right) M^2+\left(3
   M^2-10 z_\chi^2\right) A_0\left(M^2\right) \nonumber\\
& + & z_\chi^2
   \left[M^2-2 \left(M^2-4 z_\chi^2\right)
   B_0\left(z_\chi^2,0,z_\chi^2\right)\right]-\left(3 M^4-13
   z_\chi^2 M^2+8 z_\chi^4\right)
   B_0\left(z_\chi^2,M^2,z_\chi^2\right)\Biggr\}\nonumber\\
& + & \frac{g_{NR}^2 (m_\chi+z_\chi)}{64
   F^2 m_\chi z_\chi^3
   \left[(m_\chi+z_\chi)^2-M^2\right] \pi ^2}\, \Biggl\{(3
   m_\chi-z_\chi) z_\chi A_0\left(m_\chi^2\right)
   M^2+m_\chi (m_\chi+z_\chi)\nonumber\\
& \times &  \left(-3 M^2+3 m_\chi^2+4
   z_\chi^2+3 m_\chi z_\chi\right)
   A_0\left(M^2\right)+m_\chi \Biggl[M^2 (z_\chi-3 m_\chi)
   z_\chi^2+m_\chi \left(3 m_\chi^2+z_\chi^2\right)\nonumber\\
& \times &
   \left[M^2-(m_\chi+z_\chi)^2\right]
   B_0\left(z_\chi^2,0,m_\chi^2\right)+(m_\chi+z_\chi)
   \biggl[3 M^4-\left(6 m_\chi^2+3 z_\chi m_\chi+4
   z_\chi^2\right) M^2\nonumber\\
& + &  m_\chi (m_\chi+z_\chi) \left(3
   m_\chi^2+z_\chi^2\right)\biggr]
   B_0\left(z_\chi^2,M^2,m_\chi^2\right)\Biggr]\Biggr\}\nonumber\\
& + & \frac{g_{\Delta R}^2}{2592 F^2
   z_{\Delta \chi}^4 z_\chi^3 \pi ^2}\, \Biggl\{z_\chi^2 \Biggl[\left(27 z_{\Delta
\chi}^2+20
   z_\chi z_{\Delta \chi}-5 z_\chi^2\right) M^2 +  4 z_\chi
   \biggl(-18 z_{\Delta \chi}^3-13 z_\chi z_{\Delta \chi}^2 \nonumber\\
& + & 5 z_\chi^2
   z_{\Delta \chi}+20 z_\chi^3\biggr)\Biggr] M^2-2 \biggl[-54
   z_{\Delta \chi}^4-40 z_\chi z_{\Delta \chi}^3+60 z_\chi^2
   z_{\Delta \chi}^2+10 z_\chi^4 \nonumber\\
& + & M^2 \left(27 z_{\Delta \chi}^2+20 z_\chi
   z_{\Delta \chi}-5 z_\chi^2\right)\biggr]
   A_0\left(z_{\Delta \chi}^2\right) M^2 + 2 \biggl[5 z_\chi^6+20
   z_{\Delta \chi} z_\chi^5 \nonumber\\
& + & 5 \left(3 M^2-5 z_{\Delta \chi}^2\right)
   z_\chi^4-2 z_{\Delta \chi} \left(10 M^2+53 z_{\Delta \chi}^2\right)
   z_\chi^3-\left(5 M^4-33 z_{\Delta \chi}^2 M^2+55 z_{\Delta \chi}^4\right)
   z_\chi^2 \nonumber\\
& + & 20 z_{\Delta \chi} \left(M^2-z_{\Delta \chi}^2\right)^2
   z_\chi+27 \left(z_{\Delta \chi}^3-M^2 z_{\Delta
   \chi}\right)^2\biggr]
   A_0\left(M^2\right)-2 (z_{\Delta \chi}+z_\chi)^2 \nonumber\\
& \times & \left(27
   z_{\Delta \chi}^6-34 z_\chi z_{\Delta \chi}^5-41 z_\chi^2
   z_{\Delta \chi}^4+98 z_\chi^3 z_{\Delta \chi}^3-17 z_\chi^4
   z_{\Delta \chi}^2-10 z_\chi^5 z_{\Delta \chi}-5
   z_\chi^6\right)\nonumber\\
& \times & B_0\left(z_\chi^2,0,z_{\Delta \chi}^2\right) - 2
   (M-z_{\Delta \chi}-z_\chi) (M+z_{\Delta \chi}+z_\chi) \biggl[27
   z_{\Delta \chi}^6-34 z_\chi z_{\Delta \chi}^5-41 z_\chi^2
   z_{\Delta \chi}^4\nonumber\\
& + & 98 z_\chi^3 z_{\Delta \chi}^3 - 17 z_\chi^4
   z_{\Delta \chi}^2-10 z_\chi^5 z_{\Delta \chi}-5 z_\chi^6+M^4 \left(27
   z_{\Delta \chi}^2+20 z_\chi z_{\Delta \chi}-5 z_\chi^2\right)\nonumber\\
& - & 2 M^2
   \biggl(27 z_{\Delta \chi}^4-7 z_\chi z_{\Delta \chi}^3 - 50 z_\chi^2
   z_{\Delta \chi}^2+5 z_\chi^3 z_{\Delta \chi}-5
   z_\chi^4\biggr)\biggr]
   B_0\left(z_\chi^2,M^2,z_{\Delta \chi}^2\right)\Biggr\}.
\label{G_FFLoopDiagramssST}
\end{eqnarray}


\begin{thebibliography}{99}




\bibitem{Weinberg:1979kz}
S.~Weinberg,
Physica {\bf A96}, 327 (1979).

\bibitem{Gasser:1983yg}
  J.~Gasser and H.~Leutwyler,
  Annals Phys.\  {\bf 158}, 142 (1984).

\bibitem{Scherer:2009bt}
  S.~Scherer,
  Prog.\ Part.\ Nucl.\ Phys.\  {\bf 64}, 1 (2010).

\bibitem{Gasser:1987rb}
  J.~Gasser, M.~E.~Sainio, and A.~\v{S}varc,
  Nucl.\ Phys.\  {\bf B307}, 779 (1988).

\bibitem{Jenkins:1990jv}
  E.~E.~Jenkins and A.~V.~Manohar,
  Phys.\ Lett.\  B {\bf 255}, 558 (1991).

\bibitem{Tang:1996ca}
  H.~B.~Tang,
  arXiv:hep-ph/9607436.

\bibitem{Becher:1999he}
  T.~Becher and H.~Leutwyler,
  Eur.\ Phys.\ J.\  C {\bf 9}, 643 (1999).

\bibitem{Gegelia:1999gf}
  J.~Gegelia and G.~Japaridze,
  Phys.\ Rev.\  D {\bf 60}, 114038 (1999).

\bibitem{Fuchs:2003qc}
  T.~Fuchs, J.~Gegelia, G.~Japaridze, and S.~Scherer,
  Phys.\ Rev.\  D {\bf 68}, 056005 (2003).


\bibitem{Hemmert:1997ye}
  T.~R.~Hemmert, B.~R.~Holstein, and J.~Kambor,
  J.\ Phys.\ G {\bf 24}, 1831 (1998).

\bibitem{Pascalutsa:2002pi}
  V.~Pascalutsa and D.~R.~Phillips,
  Phys.\ Rev.\  C {\bf 67}, 055202 (2003).

\bibitem{Bernard:2003xf}
  V.~Bernard, T.~R.~Hemmert, and U.-G.~Mei{\ss}ner,
  Phys.\ Lett.\  B {\bf 565}, 137 (2003).

\bibitem{Pascalutsa:2006up}
  V.~Pascalutsa, M.~Vanderhaeghen, and S.~N.~Yang,
  Phys.\ Rept.\  {\bf 437}, 125 (2007).

\bibitem{Hacker:2005fh}
  C.~Hacker, N.~Wies, J.~Gegelia, and S.~Scherer,
  Phys.\ Rev.\  C {\bf 72}, 055203 (2005).

\bibitem{Stuart:1990}
R.~G.~Stuart, in ${\rm Z}^0$ {\it Physics}, ed. J. Tran Thanh Van
(Editions Frontieres, Gif-sur-Yvette, 1990), p.\ 41.

\bibitem{Denner:1999gp}
  A.~Denner, S.~Dittmaier, M.~Roth, and D.~Wackeroth,
  Nucl.\ Phys.\  {\bf B560}, 33 (1999).

\bibitem{Denner:2006ic}
  A.~Denner and S.~Dittmaier,
  Nucl.\ Phys.\ Proc.\ Suppl.\  {\bf 160}, 22 (2006).

\bibitem{Actis:2006rc}
  S.~Actis and G.~Passarino,
  Nucl.\ Phys.\ {\bf B777}, 100 (2007).

\bibitem{Actis:2008uh}
  S.~Actis, G.~Passarino, C.~Sturm, and S.~Uccirati,
  Phys.\ Lett.\ B {\bf 669}, 62 (2008).

\bibitem{Djukanovic:2009zn}
  D.~Djukanovic, J.~Gegelia, A.~Keller, and S.~Scherer,
  Phys.\ Lett.\  B {\bf 680}, 235 (2009).

\bibitem{Djukanovic:2009gt}
  D.~Djukanovic, J.~Gegelia, and S.~Scherer,
  Phys.\ Lett.\  B {\bf 690}, 123 (2010).


\bibitem{Rarita:1941mf}
W.~Rarita and J.~S.~Schwinger,
Phys.\ Rev.\  {\bf 60}, 61 (1941).

\bibitem{Borasoy:2006fk}
  B.~Borasoy, P.~C.~Bruns, U.-G.~Mei{\ss}ner, and R.~Lewis,
  Phys.\ Lett.\  B {\bf 641}, 294 (2006).

\bibitem{Cutkosky:1960sp}
  R.~E.~Cutkosky,
  J.\ Math.\ Phys.\  {\bf 1}, 429 (1960).

\bibitem{BGJS}
T.~Bauer, J.~Gegelia, G.~Japaridze, and S.~Scherer, in preparation.

\bibitem{Veltman:1963th}
  M.~J.~G.~Veltman,
  Physica {\bf 29}, 186 (1963).

\bibitem{Weinberg:1991um}
  S.~Weinberg,
  Nucl.\ Phys.\  B {\bf 363}, 3 (1991).

\bibitem{Ecker:1995gg}
  G.~Ecker,
  Prog.\ Part.\ Nucl.\ Phys.\  {\bf 35}, 1 (1995).


\bibitem{Djukanovic:2007bw}
  D.~Djukanovic, J.~Gegelia, and S.~Scherer,
  Phys.\ Rev.\  D {\bf 76}, 037501 (2007).

\bibitem{Gegelia:2009py}
  J.~Gegelia and S.~Scherer,
  Eur.\ Phys.\ J.\  A {\bf 44}, 425 (2010).

\bibitem{Arndt:1985vj}
  R.~A.~Arndt, J.~M.~Ford and L.~D.~Roper,
  Phys.\ Rev.\  D {\bf 32}, 1085 (1985).

\bibitem{Cutkosky:1990zh}
  R.~E.~Cutkosky and S.~Wang,
  Phys.\ Rev.\  D {\bf 42}, 235 (1990).


\bibitem{Bincer:1959tz}
  A.~M.~Bincer,
  Phys.\ Rev.\  {\bf 118}, 855 (1960).

\bibitem{Naus:1987kv}
  H.~W.~L.~Naus and J.~H.~Koch,
  Phys.\ Rev.\ C {\bf 36}, 2459 (1987).

\bibitem{Koch:2001ii}
  J.~H.~Koch, V.~Pascalutsa, and S.~Scherer,
  Phys.\ Rev.\ C {\bf 65}, 045202 (2002).

\bibitem{Ward:1950xp}
  J.~C.~Ward,
  Phys.\ Rev.\  {\bf 78}, 182 (1950).

\bibitem{Takahashi:1957xn}
  Y.~Takahashi,
  Nuovo Cim.\  {\bf 6}, 371 (1957).

\bibitem{Fuchs:2003ir}
  T.~Fuchs, J.~Gegelia, and S.~Scherer,
  J.\ Phys.\ G {\bf 30}, 1407 (2004).

\bibitem{PDG}
  K.~Nakamura {\it et al.} [ Particle Data Group Collaboration ],
  J.\ Phys.\ G {\bf 37}, 075021 (2010).

\bibitem{GellMann:1968rz}
  M.~Gell-Mann, R.~J.~Oakes, B.~Renner,
  Phys.\ Rev.\  {\bf 175}, 2195-2199 (1968).

\bibitem{Mertig:1990an}
  R.~Mertig, M.~Bohm, and A.~Denner,
  Comput.\ Phys.\ Commun.\  {\bf 64}, 345 (1991).


\bibitem{Denner:2005nn}
  A.~Denner and S.~Dittmaier,
  Nucl.\ Phys.\  B {\bf 734}, 62 (2006).

\end{thebibliography}
\end{document}